\documentstyle[editedbook,epsfig,psfig,epsf]{mq}
\def\cm2{cm$^2$ }
\def\se1{s$^{-1}$ }

\newcommand{\ee}{$e^\pm$}
\newcommand{\g}{$\gamma$}

\newcommand{\sax}{{\it Beppo\-SAX}}

\newcommand{\xte}{{\it RXTE}}

\newcommand{\gro}{{\it CGRO}}

\newcommand{\aap}{{A\&A}}
\newcommand{\aaps}{{A\&AS}}
\newcommand{\mnras}{{MNRAS}}
\newcommand{\apj}{{ApJ}}

\begin{opening}
\title{Radiative Processes in Microquasars}
\author{Juri Poutanen$^1$ \& Andrzej A. Zdziarski$^{2}$}
\institute{$^1$ Astronomy Division, P.O.Box 3000, 90014 University of Oulu, Finland \\
$^2$ Centrum Astronomiczne im.\ M. Kopernika, Bartycka 18, 00-716 Warszawa, Poland}
\end{opening}

\runningtitle{Radiative Processes in Microquasars}
\runningauthor{Poutanen \& Zdziarski}

\begin{document}
\vspace{-0.5cm}
\begin{abstract}
{\small 
Recent advances in the X-ray and soft $\gamma$-ray 
observations of accreting black holes and microquasars, in 
particular, are reviewed. The radiative processes responsible for the emission 
are discussed briefly. The hybrid thermal/nonthermal Comptonization model 
is shown to describe well the observed broad-band spectra.  
We also comments on alternative phenomenological 
and physical models that are used to describe the X/$\gamma$-ray spectra of accreting 
black holes. Among those are the ``standard'' model (i.e. disk-blackbody plus a power-law), 
{\sc pexrav}, bulk motion Comptonization, and 
synchrotron emission from the jet. 
}
\end{abstract}

\section{Introduction}

\begin{figure}[htb]
\centerline{\epsfig{file=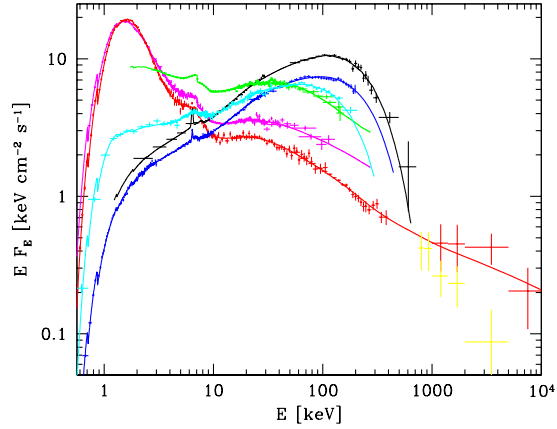,width=7.0cm}}
\vspace{-0.2cm}
\caption{A collection of broad-band spectra of Cyg X-1. The solid curves
give the best-fit Comptonization models (thermal in the hard state, and hybrid,
thermal-nonthermal in the other states). 
From \protect\cite{aaz02}. }
\label{fig:cygx1}
\vspace{-0.3cm}
\end{figure}

Accreting black holes radiate in the two main spectral states 
which we refer to later as hard and soft (see Fig.~\ref{fig:cygx1}). 
The hard state is characterized by a power-law--like spectrum which abruptly 
cuts off  at $\sim$ 100 keV. The energy output is 
dominated by the 100 keV photons \cite{aaz00}. 
Such a spectrum is
interpreted as Comptonization by thermal electrons in the inner 
hot disk or active magnetic corona above the accretion disk (e.g. \cite{pkr97,b99,mal01}). 
A weak MeV tail observed in Cyg X-1  is probably 
a signature of non-thermal electrons in the source \cite{li96,pc98,mc02}.

In the soft state the emission is dominated 
by the black-body--like component peaking at a few keV 
with a power-law--like component above 30 keV extending 
up to 1 MeV or even higher \cite{aaz00,mc02,p98,grove98}. 
These spectra cannot be fitted with thermal Comptonization models and 
require  the radiating electrons to have a significant non-thermal fraction  
\cite{pc98,gier99,aaz01}.

\section{Spectral Modeling}

\subsection{Cygnus X-1}

\begin{figure}%[htb]
\centerline{\epsfig{file=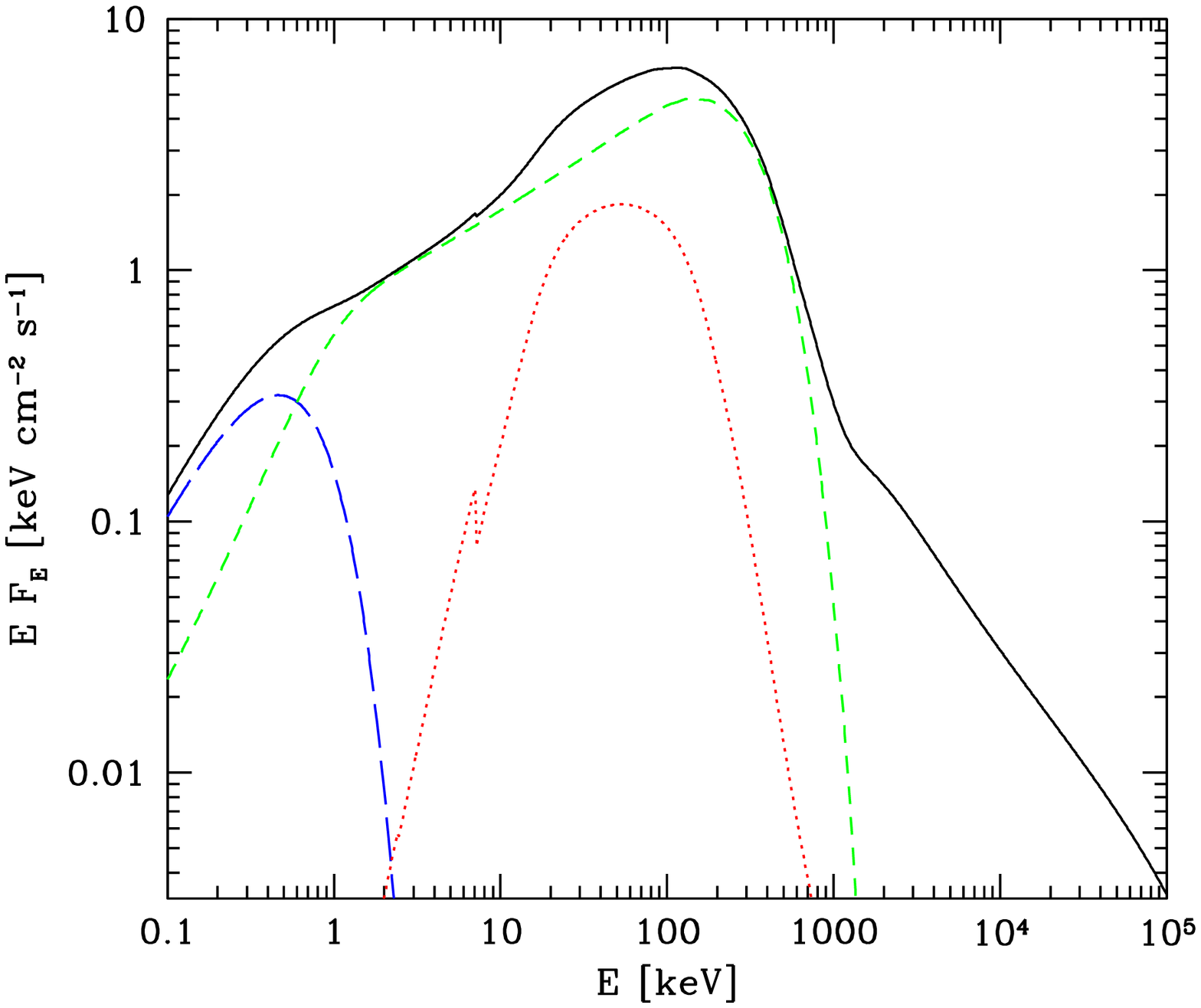,width=6.5cm}
\epsfig{file=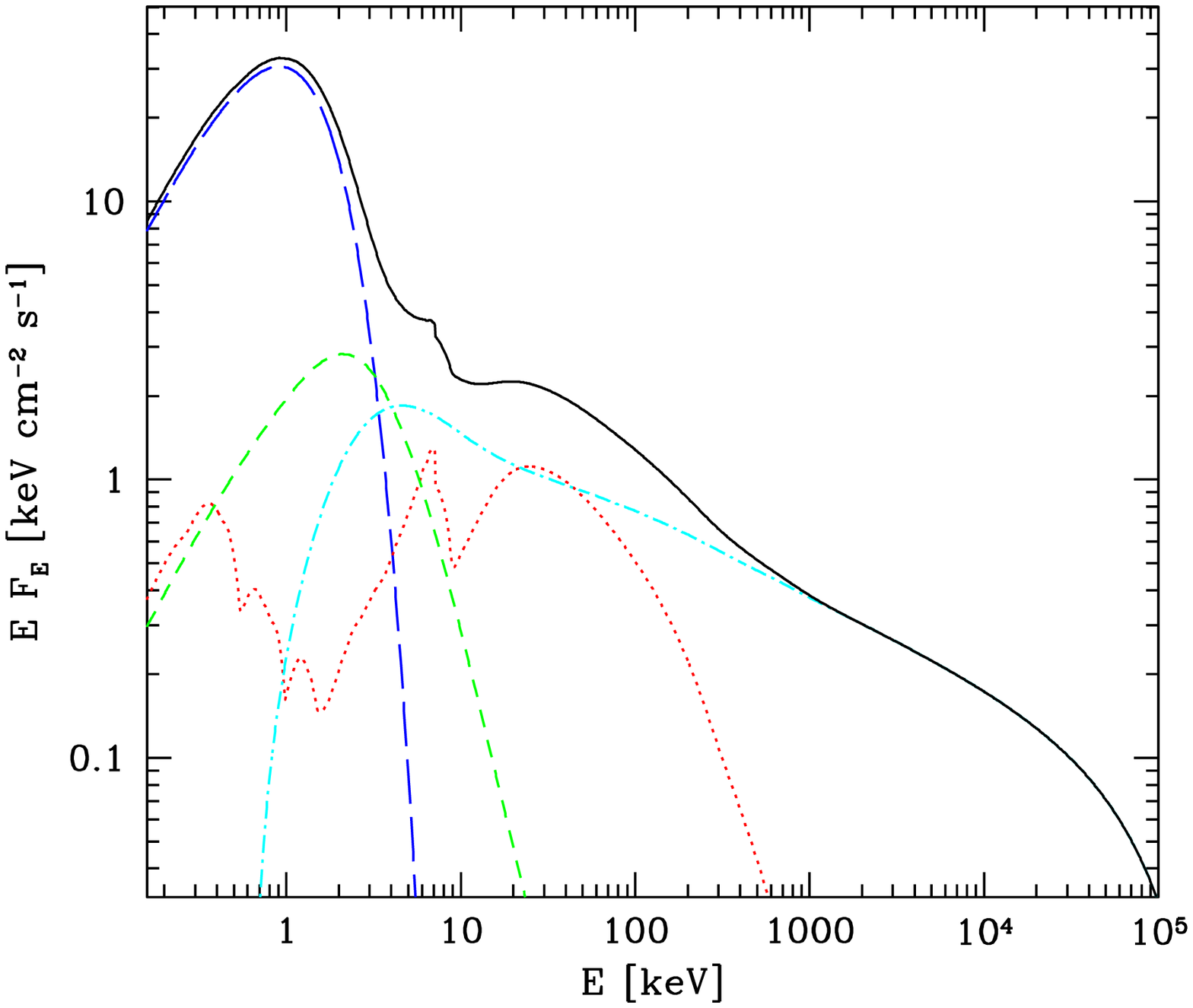,width=6.4cm}}
\vspace{-0.2cm}
\caption{Left: Components of the {\sc eqpair} fit to the hard state \gro\/
data of Cyg X-1.
 The long dashes, short dashes,
and dots correspond to the unscattered blackbody, scattering by
thermal electrons, and Compton reflection, respectively. The solid curve is the
total spectrum. Scattering by the nonthermal electrons accounts for the
high-energy tail above the thermal-Compton spectrum given by the short
dashes, starting at $\sim 1$ MeV. \protect\\
Right:  Components of the {\sc eqpair} fit to the \sax-\gro\/ data
for the soft state of Cyg X-1. 
 The curves have the same meaning as in the left panel. The 
dots/dashes  correspond to the scattering by nonthermal electrons.
All spectra are intrinsic, i.e., corrected for absorption.
From \protect\cite{mc02}. }
\label{fig:components}
\vspace{-0.3cm}
\end{figure}

A weak MeV tail observed in Cyg X-1 in its hard state already shows us 
that the emitting electrons cannot be purely thermal 
(see left panel of Fig.~\ref{fig:components}). 
An obvious generalization is to assume that there is a 
non-thermal tail in the electron distribution that is produced
by some acceleration process (e.g. \cite{li96}). 
The soft state data (even below 1 MeV) could not be fitted at all with thermal 
models. Modeling of the spectral transitions with a generalized hybrid thermal/non-thermal 
model {\sc eqpair} \cite{pc98,c99} 
predicted a stronger power-law--like tail in the soft state 
extending up to 10 MeV. \gro/COMPTEL observations  confirmed 
the existence of this tail: the decrease of the hard X-ray luminosity was 
accompanied by the {\it increase} of the soft $\gamma$-ray luminosity \cite{mc02}. 
In the context of the hybrid Comptonization model, the 
power-law is a result of single Compton scattering off non-thermal 
population of electrons  (see right panel in Fig.~\ref{fig:components}). 
A cutoff at about 10 MeV (depending on the compactness of the source) 
should appear in the spectrum due to the
absorption of the $\gamma$-rays by softer photons resulting in pair 
production.

\subsection{GRS 1915+105}

\begin{figure}%[htb]
\centerline{\epsfig{file=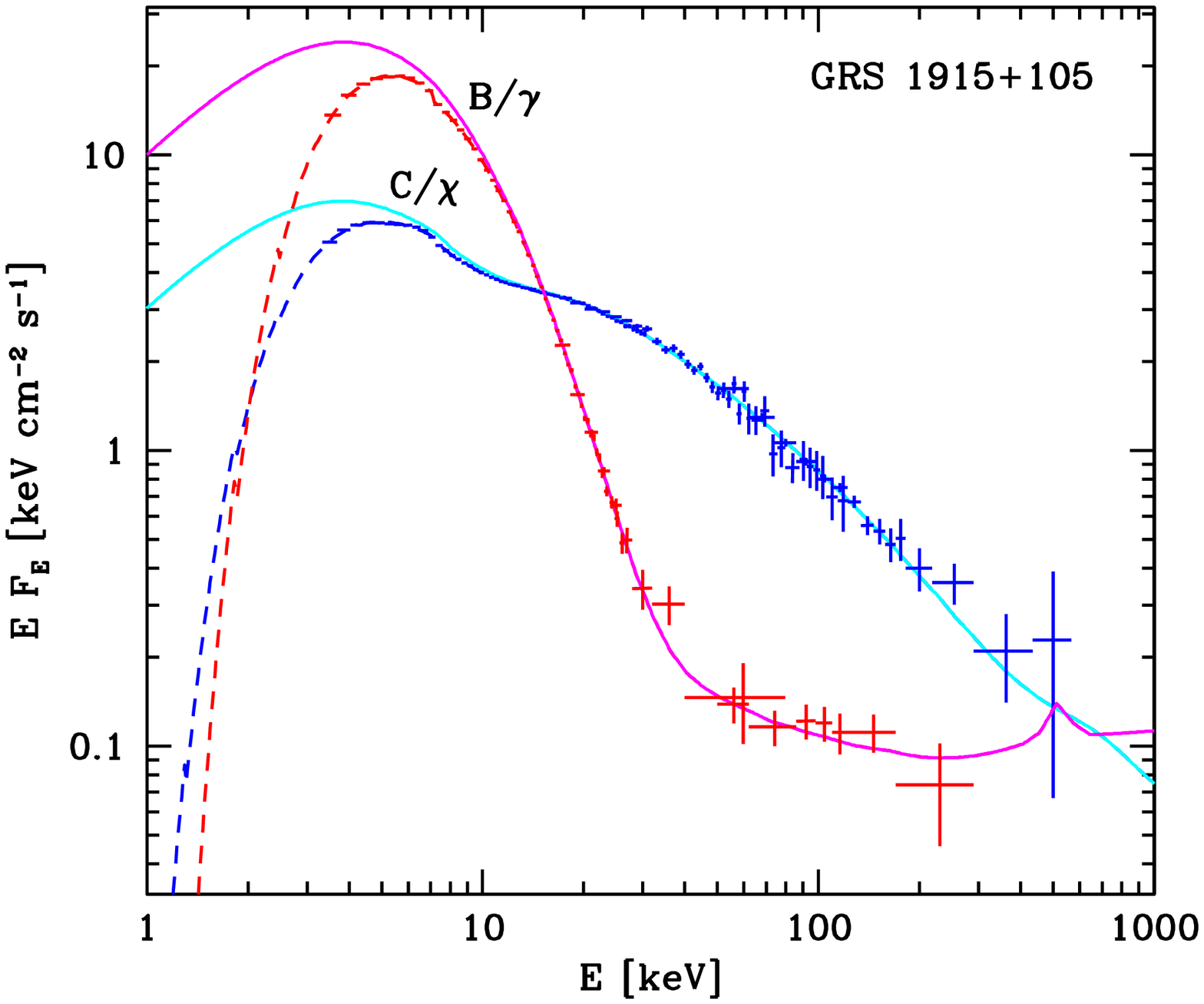,width=7.0cm}
\epsfig{file=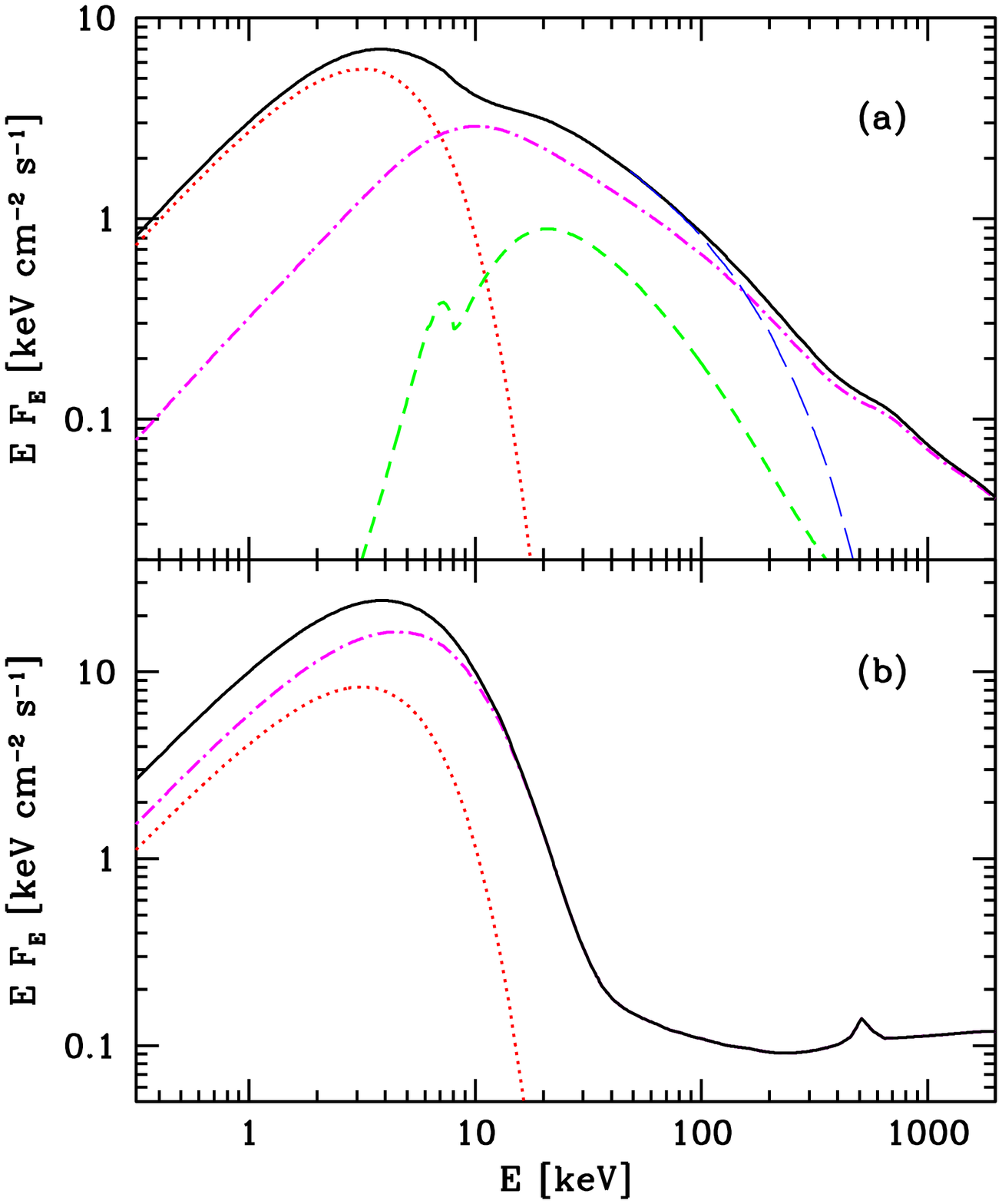,width=6cm}}
\caption{Left: Fits to simultaneous {\it RXTE}-OSSE spectra of GRS 1915+105 from VP 619 (1997 
May 14--20) and VP 813 (1999 April 21--27)
with the hybrid Comptonization model {\sc eqpair}. The dashed and solid curves show the
models of the observed spectra and the unabsorbed spectra, respectively. \protect\\
Right: (a) Components of the fit to the VP 619 data. All spectra are
intrinsic, i.e., corrected for absorption. The dotted, dot-dashed and dashed
curves give the unscattered blackbody component, the scattered spectrum, and
the component due to Compton reflection and Fe K$\alpha$ fluorescence,
respectively. The solid curve is the total spectrum. The thin long-dashed curve
shows the best-fit thermal Comptonization model, which lies much below the
data above 100 keV. (b) The total model spectrum and the corresponding two
components for the VP 813 data. The cutoff at $\sim 10$ MeV is due to 
photon-photon pair production absorption. From \protect\cite{aaz01}.}
\label{fig:grs1915}
\vspace{-0.3cm}
\end{figure}

In spite of the fact that microquasar GRS 1915+105 show dramatic 
variability pattern, almost all its hard X-ray spectra are remarkably 
similar. During eight (out of nine) observations with the \gro/OSSE  
the source showed a simple power-law--like spectrum in the 50--500 keV band 
with photon index $\Gamma\approx 3$ \cite{aaz01}. Only in one occasion (when the X-ray flux 
was very high), the hard X-ray spectrum was much harder $\Gamma\approx 2.3$ and the flux was low.
We note here that in all observations the source has a much softer 
spectrum than the normal hard state of  Cyg X-1, i.e. it was always in the 
soft state. There is no signature of the high-energy cutoff in the data.
The {\sc eqpair} model gives a good description of the data 
(see Fig.~\ref{fig:grs1915}) indicating that  
about 10-20\% of the total power goes to accelerate non-thermal electrons. 
The  C/$\chi$-state   \cite{bel00}  differs, however, from the  B/$\gamma$-state
in that the 20-200 keV tail is produced by thermal Comptonization in 
the former and by non-thermal Compton scattering in the later.

\section{Old and New Alternatives}

\subsection{The ``standard'' model}

The black body looking soft component is associated with the optically thick 
accretion disk by most of the researchers. The broad-band 
spectra are often fitted by the ``standard'' model consisting of a  disk-blackbody (soft component) 
and a power-law (hard). There are numerous problems with such modeling. 
First, a black body is a bad representation of the  spectrum expected
from the accretion disk (e.g. \cite{mer00}). Real data also show that 
the soft bumps in the Cyg X-1 soft state \cite{gier99} and GRS 1915+105  \cite{aaz01} cannot be 
fitted by a black body (or multicolor disk). Thermal Comptonization  of a blackbody
is a much better description of these spectra. 
Second, a power-law, even exponentially folded, is a very bad 
representation of the Comptonization spectra. 
At the lower end, Comptonization spectrum cuts off below the seed photon energy 
while a power-law has no break there. 
The normalization of the blackbody thus can be underestimated by a large 
factor   (see e.g. \cite{vilhu01}). 
The conclusions (e.g. variations of the inner disk radius) 
resulting from fitting the data with 
this  ``standard'' model thus should be taken with a grain of salt.

At the higher end, the (thermal) Comptonization spectrum has a 
much sharper cutoff than an exponentially folded power-law. 
This difference in the spectral shape is important 
when we model the broad-band spectra from accreting black holes with 
the later model adding a Compton reflection component (model {\sc pexrav} 
\cite{mz95} from XSPEC) since the amplitude of the reflection component
strongly depends on the assumed shape of the underlying continuum.
Thus, we would advise not to use {\sc pexrav}
when modeling Comptonization spectra close to the black body 
or to the high energy cutoff. 

\subsection{X-rays from the jet?}

A very interesting correlation between radio and X-ray fluxes has been 
discovered recently \cite{bro99,cor00,gal02}. There are two possible origins of this 
correlation. One is that the level of X-ray emission is related to the rate of 
ejection of radio-emitting clouds, forming a compact jet (e.g., 
\cite{cor00,mir98}). Another is that the X-ray emission of black hole binaries 
is dominated by the synchrotron   emission of the jet \cite{mar01,mar02}.

We note here there are many strong arguments against the second interpretation. 
The broad-band X/\g-ray spectra of black hole binaries in the hard state are 
very well modeled by thermal Comptonization and Compton reflection (e.g., 
\cite{gier97,aaz98,f01a,f01b}).  The presence of reflection implies that the 
X-ray emission is not strongly beamed away from the disk. The thermal-Compton 
origin of the primary X-ray emission is strongly supported by a remarkable 
uniformity of the both energy and shape of the high-energy cutoffs of black hole 
binaries in the hard state observed by OSSE \cite{grove98}. This cutoff is 
naturally accounted for by thermostatic properties of thermal Comptonization as 
well as \ee\ pair production (e.g., \cite{mal01}), as it corresponds to the 
transition to relativistic temperatures. At higher temperatures, cooling becomes 
extremely efficient and copious pair production starts. This reduces the energy 
available per particle leading to the temperature decrease.  On the other hand, 
$m_{\rm e} c^2$ plays no particular role for non-thermal synchrotron 
emission (cf. variable cutoff energy during flares in blazars). Thus, accounting for the 
observed cutoff energies requires fine-tuning of product of 
the square of the maximum electron energy and the magnetic field strength 
in the non-thermal synchrotron  models. 
In addition, the jet model has problems reproducing the 
actual shape of the cutoff. For example, the synchrotron model of the hard state 
of Cyg X-1  (see fig. 3a in \cite{mar02}) when matched to the 100 keV flux 
overestimates the 1 MeV flux (see \cite{mc02}) by a factor of 8.

\begin{figure}%[htb]
\centerline{\epsfig{file=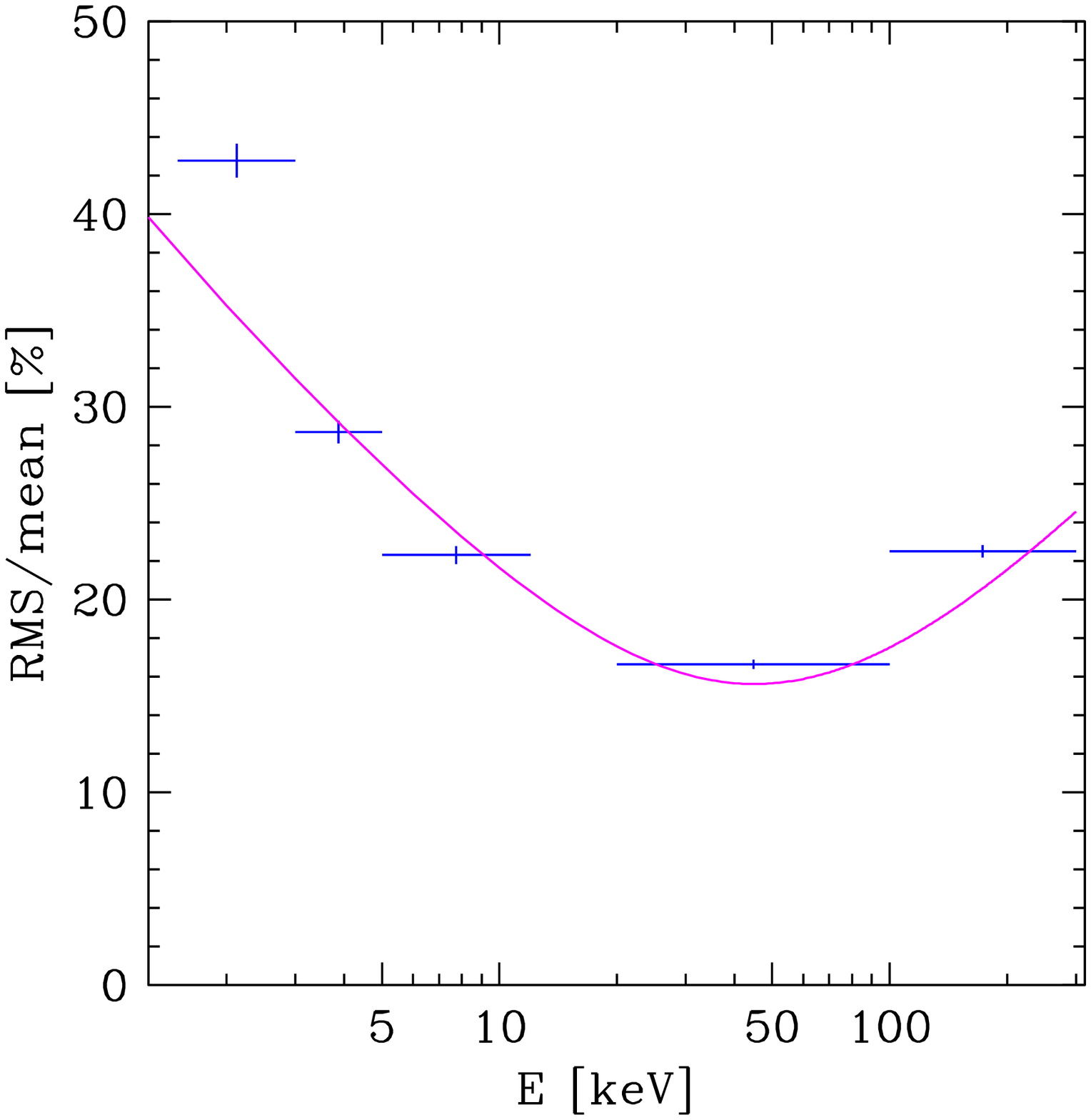,width=6.0cm}
\hspace{0.5cm}
\epsfig{file=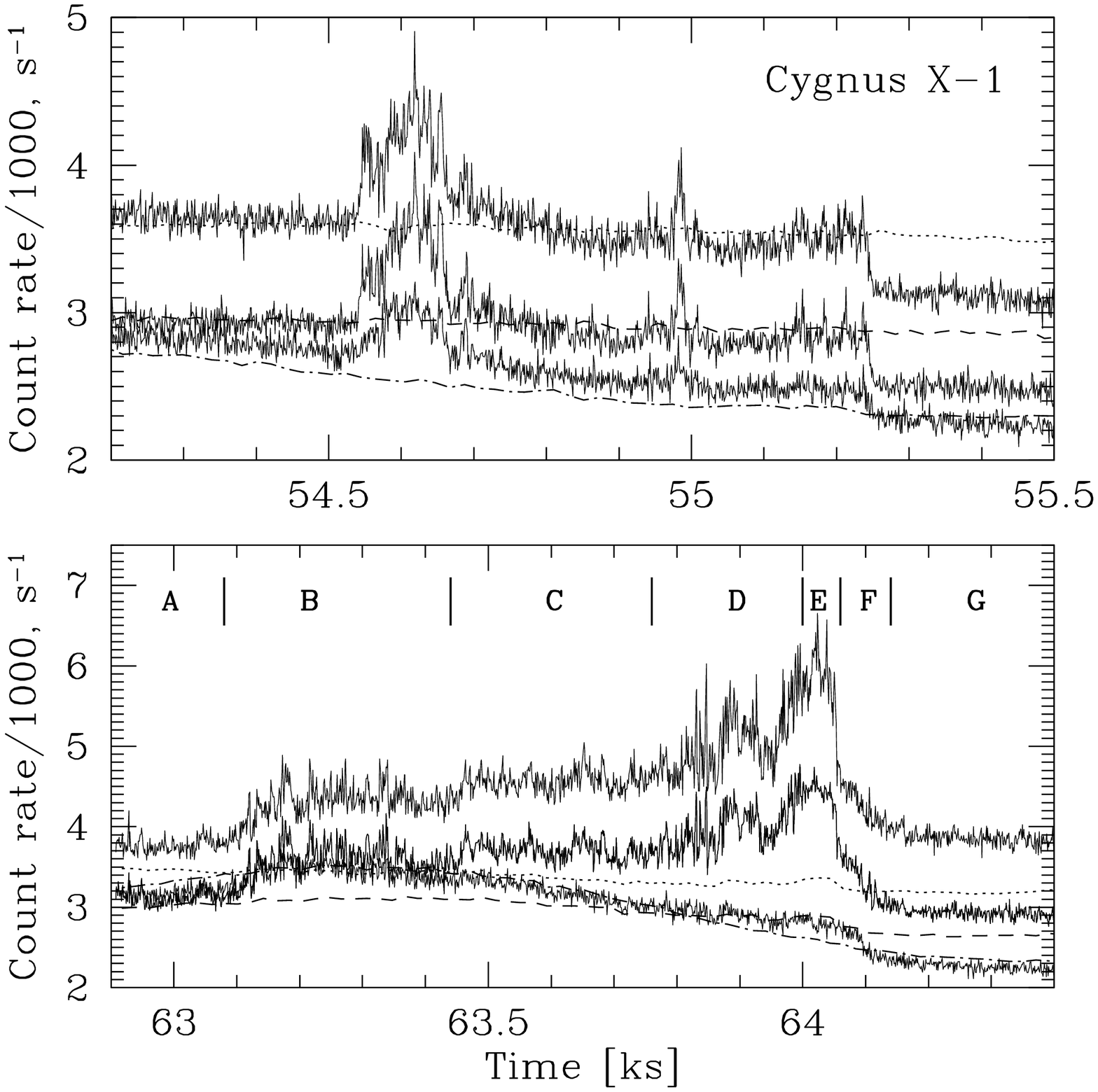,width=6.5cm}}
\caption{Left: The rms variability in the hard state of Cyg X-1 in one-day averaged
data from the \xte/ASM and \gro/BATSE \protect\cite{aaz02}. The model curve 
corresponds to a power-law pivoting with $\Delta\Gamma\simeq 0.2$ 
around the energy which has a Gaussian distribution 
around 45 keV (see \protect\cite{aaz002} for details).  \protect\\
Right: Count rate during the two outbursts of Cyg X-1 on 1999 April 21 
in the BATSE large area detectors energy  channels 1--3 (corresponding to 
approximately 20--50 keV, 50--100 keV, 100--300 keV). Count rates are higher
in  softer  channels. 
The count rate is summed over two
detectors  closest to the line of sight to Cyg~X-1.    
Dashes, dots, and dash/dots show  the background in channels 1, 2, 3,  as seen by two detectors
looking away from Cyg~X-1. From \protect\cite{stern01}. 
}
\label{fig:rms}
\vspace{-0.5cm}
\end{figure}

An additional evidence against a substantial part of X-rays being non-thermal 
synchrotron is provided by spectral variability. In the case of Cyg X-1, the 
ASM/BATSE data show spectral pivoting around $\sim 40$--50 keV (see \cite{aaz02} 
and Fig.~\ref{fig:rms}). The characteristic  variations of the power-law slope 
$\Delta \Gamma$ from those data is $\sim 0.2$--0.3. This power-law spectral 
variability extended to 15 GHz would imply a huge variability of the radio flux, 
by several orders of magnitude. However, the range of the variability of the 15 
GHz flux correlated with the ASM flux is by a factor of several, basically the 
same as the range of the variability of the ASM flux itself \cite{gal02}. If 
both radio and X-rays were due to non-thermal synchrotron emission, their 
observed variability pattern should yield the rms in X-rays virtually 
independent of energy. This is clearly in strong disagreement with the data 
shown in Fig.\ \ref{fig:rms}, and, in particular, with the ASM 1.5--3 keV flux 
being strongly anticorrelated with the 100--300 keV flux from BATSE \cite{aaz02}. 

The amplitude of Compton reflection and the iron line imply that  
dense and rather cold material occupies a solid angle $\sim\pi$ 
as viewed from the X-ray source. Smearing of these components 
and the observed correlations with the spectral slope \cite{aaz002,aaz99,gil00}  
clearly identifies the reflector with the accretion disk and implies the origin 
of the continuum emission within 30--100 gravitational radii 
from the black hole which would also be consistent with 
rapid X-ray variability.  
The above arguments also rule out the dominant contribution of the 
nonthermal Compton emission \cite{geo02} from the jet to the observed hard state 
spectra of black holes.

All these arguments strongly support the interpretation of the correlated radio emission as
being due to ejection of clouds from the X-ray source  
which could be similar to coronal mass ejections (CME) observed at the Sun.
The recently discovered strong  X/$\gamma$-rays flares from Cyg X-1 
\cite{stern01,gol02} could be the extremes of such an activity. 
The right panel in Fig.~\ref{fig:rms} shows the flaring activity of Cyg X-1 
in April 1999. The  episode D--E  shows strong flare in the 20--100 keV band with a weaker 
activity above 100 keV and no detectable signal above 300 keV.
This indicates that there exist at least two independent 
spectral components (see \cite{stern01}): 
one could be related to the inner hot disk 
or the magnetized corona, while another to  the base of the jet.

\subsection{Bulk motion Comptonization} 

The power-law like spectra of black holes in the soft state 
were interpreted as resulting from  bulk motion Comptonization 
in the converging flow \cite{tit98,bor99,tit99}. The  
specific feature of that model is a cutoff at 
$\sim 100$--200 keV. (We note here that the XSPEC version of the model {\sc bmc}
has no cutoff built in.) 
The data show no signatures of the cutoff at least up to 500 keV in GRS 1915+105 
\cite{aaz01}  and up to 10 MeV in Cyg X-1  \cite{mc02}.
This supports their non-thermal origin and  strongly rules 
out a significant contribution from the bulk motion Comptonization.
(See further discussion in \cite{aaz00,aaz01}.)

\section*{Acknowledgments}
This work was partly supported by the Academy of Finland and grants from the 
Polish Committee for Scientific Research (5P03D00821, 2P03C00619p1,2) and the 
Foundation for Polish Science.


\begin{thebibliography}{}

\bibitem{aaz02} Zdziarski A. A., Poutanen J., Paciesas W. S., \& Wen L., 
2002, \apj, {\bf 578}, in press (astro-ph/0204135).

\bibitem{aaz00}
Zdziarski A. A., 2000, in IAU Symp.\ 195, Highly Energetic Physical Processes
and Mechanisms for Emission from Astrophysical Plasmas, eds.\ P. C. H. Martens,
S. Tsuruta \& M. A. Weber (San Francisco: ASP), 153 (astro-ph/0001078).

\bibitem{pkr97} Poutanen J., Krolik J. H., \& Ryde F.,  1997, \mnras, {\bf 292}, L21.

\bibitem{b99} Beloborodov A. M., 1999, \apj,  {\bf 510}, L123.

\bibitem{mal01} Malzac J., Beloborodov A. M., \& Poutanen J., 2001, \mnras, {\bf 326}, 417

\bibitem{li96} Li H., Kusunose M., \&  Liang E. P., 1996, \aaps, {\bf 120C}, 167. 

\bibitem{pc98} Poutanen J., \& Coppi P. S., 1998, Physica Scripta, {\bf T77}, 57.

\bibitem{mc02} McConnell M. L., et al., 2002, \apj, {\bf 572}, 984.

\bibitem{p98} Poutanen J., 1998, in Theory of Black
Hole Accretion Disks, eds. M. A. Abramowicz, G. Bj\"ornsson \& J. E. Pringle
(Cambridge Univ. Press: New York), p. 100.

\bibitem{grove98} Grove J. E., et al.,  1998, \apj, {\bf 500}, 899.

\bibitem{gier99} Gierli\'nski M., Zdziarski A. A., Poutanen J., Coppi P. S., Ebisawa K., 
\& Johnson W. N., 1999, \mnras, {\bf 309}, 496.

\bibitem{aaz01} Zdziarski A. A., Grove J. E., Poutanen J., Rao A. R., \& Vadawale S. V., 
2001, \apj, {\bf 554}, L45.

\bibitem{c99} Coppi P. S., 1999, in ASP Conf.\ Ser.\ Vol.\ 161, High
Energy Processes in Accreting Black Holes, eds.\ J. Poutanen \& R. Svensson
(San Francisco: ASP), 375.

\bibitem{bel00} Belloni T., et al., 2000, \aap,  {\bf 355}, 271.

\bibitem{mer00} Merloni A., Fabian A. C., \& Ross R. R., 2000, \mnras, {\bf 313}, 193. 

\bibitem{vilhu01} Vilhu O., Poutanen J., Nikula P., \& Nevalainen J., 2001, \apj, {\bf 553}, L51. 

\bibitem{mz95} Magdziarz P., \& Zdziarski A. A., 1995, \mnras, 273, 837.

\bibitem{bro99} Brocksopp C., et al., 1999, \mnras, {\bf 309}, 1063.

\bibitem{cor00} Corbel S., et al., 2000, A\&A, {\bf 359}, 251.

\bibitem{gal02} Gallo E., Fender R., \& Pooley G. G., 2002, these proceedings.

\bibitem{mir98} Mirabel I.~F., et al. 1998, A\&A, {\bf 330}, L9. 

\bibitem{mar01} Markoff S., Falcke H., \& Fender R., 2001, \aap, {\bf 372}, L25.

\bibitem{mar02} Markoff S., Nowak M., Corbel S., Fender R., Falcke H., 2002, these proceedings.

\bibitem{gier97}
Gierli\'nski M., Zdziarski A. A, Done C., Johnson W. N., Ebisawa K., Ueda Y.,
Haardt F., \& Phlips B. F., 1997,  \mnras, {\bf 288}, 958

\bibitem{aaz98}
Zdziarski A. A., Poutanen J., Miko{\l}ajewska J., Gierli\'nski M., Ebisawa
K., \& Johnson W. N., 1998, \mnras, {\bf 301}, 435.

\bibitem{f01a} Frontera F. et al., 2001a,  \apj,{\bf  546}, 1027.

\bibitem{f01b} Frontera F. et al., 2001b, \apj, {\bf  561}, 1006.

\bibitem{aaz002} Zdziarski A. A., Gilfanov M., Lubi\'nski P., \& Revnivtsev M., 2002, in 
preparation. 

\bibitem{aaz99} Zdziarski A.  A., Lubi\'nski P., \& Smith D. A., 1999, \mnras, {\bf 303}, L11.

\bibitem{gil00} Gilfanov M., Churazov E., \& Revnivtsev M., 2000,
in Zhao G., Wang J. J., Qiu H. M., Boerner G., eds,
SGSC Conference Series Vol. 1,
5th Sino-German Workshop on Astrophysics.
China Science \& Technology Press, Beijing, 114
(astro-ph/0002415).

\bibitem{geo02} Georganopoulos M., Aharonian F.~A., \& Kirk J.~G., 2002, A\&A, {\bf 388}, L25.

\bibitem{stern01} Stern B. E., Beloborodov A. M., \& Poutanen J.,  2001, \apj, {\bf 555}, 829.

\bibitem{gol02} Golenetskii S., et al., 2002, IAUC 7840.

\bibitem{tit98} Shrader C., \& Titarchuk L., 1998, ApJ, {\bf 499}, L31.

\bibitem{bor99} Borozdin K., Revnivtsev M., Trudolyubov S., Shrader C., \& Titarchuk L., 
1999, \apj, {\bf 517}, 367. 

\bibitem{tit99} Laurent P., \& Titarchuk L., 1999, ApJ, {\bf 511}, 289.


\end{thebibliography}
\end{document}